\documentclass[a4paper]{article}

\usepackage{INTERSPEECH2022}
\usepackage{float}
\usepackage{hyperref}
\usepackage{amsmath}
\usepackage{booktabs}
\usepackage{diagbox}
\usepackage{multirow}
\usepackage{color}
\usepackage{graphicx}
\usepackage{subfigure}

\title{Learn2Sing 2.0: Diffusion and  Mutual Information-Based Target Speaker SVS by Learning from Singing Teacher}
\name{Heyang Xue$^{1,3}$, Xinsheng Wang$^2$, Yongmao Zhang$^2$, Lei Xie$^{1,2,*}$, Pengcheng Zhu$^3$, Mengxiao Bi$^3$}
\address{Audio, Speech and Language Processing Group (ASLP@NPU),
$^1$School of Software,\\ $^2$School of Computer Science, Northwestern Polytechnical University, Xian, China \\
$^3$Fuxi AI Lab, NetEase Inc., Hangzhou, China}
\email{xueheyang@corp.netease.com, w.xinshawn@gmail.com, zym@mail.nwpu.edu.cn,  lxie@nwpu.edu.cn, zhupengcheng@corp.netease.com, bimengxiao@corp.netease.com}

\begin{document}

\maketitle
\begin{abstract}
\vspace{-0.1cm}
Building a high-quality singing corpus for a person who is not good at singing is non-trivial, thus making it challenging to create a singing voice synthesizer for this person. Learn2Sing is dedicated to synthesizing the singing voice of a speaker without his or her singing data by learning from data recorded by others, i.e., the singing teacher. Inspired by the fact that pitch is the key style factor to distinguish singing from speaking voice, the proposed Learn2Sing~2.0 first generates the preliminary acoustic feature with averaged pitch value in the phone level, which allows the training of this process for different styles, i.e., speaking or singing, share same conditions except for the speaker information. Then, conditioned on the specific style,  a diffusion decoder, which is accelerated by a fast sampling algorithm during the inference stage, is adopted to gradually restore the final acoustic feature. During the training, to avoid the information confusion of the speaker embedding and the style embedding, mutual information is employed to restrain the learning of speaker embedding and style embedding. Experiments show that the proposed approach is capable of synthesizing high-quality singing voice for the target speaker without singing data with 10 decoding steps. \footnote{Our code and audio samples are available at \url{https://welkinyang.github.io/Learn2Sing2.0/}.}
\end{abstract}
\renewcommand{\thefootnote}{\fnsymbol{footnote}}
\footnotetext{Corresponding author.}

\vspace{-0.3cm}
\section{Introduction}

Singing voice synthesis~(SVS), which aims at generating the highly expressive singing voice, has attracted a lot of attention in recent years because of its various potential application scenarios. A typical way to perform the SVS is to generate acoustic features, i.e., mel-spectrogram~\cite{DBLP:conf/iscslp/GuYRWTZCWM21, DBLP:conf/interspeech/LeeKBLKC21}, or directly produce waveform~\cite{DBLP:journals/corr/abs-2110-08813} with lyrics and musical score as input. To this end, a singing corpus recorded by a professional singer and labeled with MusicXML or MIDI files is necessary to train a model for the SVS task. However, due to the challenge to create such a corpus for a person who is not good at singing, it is non-trivial to build a singing voice synthesizer in a typical way. This paper focuses on synthesizing high-quality singing voice for a target speaker without singing training data.

While some openly accessible singing voice databases have been released recently with the aim to promote the development of SVS~\cite{DBLP:journals/corr/abs-2201-07429, DBLP:conf/mm/HuangC0LCZ21, DBLP:journals/speech/SharmaGVTL21}, generating singing voice for a specific target speaker is still a challenge due to the difficulty of creating a singing corpus for a person who is not good at singing. Such difficulties also hinder the further promotion and practical application of SVS technology. Compared with 
singing voice data, speech data is easier to collect for a person, thus making it an interesting topic to obtain an SVS system for a person only with speech corpus by learning from another singing corpus. Most recently, several efforts have been conducted towards singing voice generation in the absence of singing voice data from the target speaker~\cite{DBLP:conf/icassp/ValleLPC20,DBLP:conf/slt/XueYLXL21,DBLP:conf/asru/LiuWLSS21}.

As the first study towards this task, Mellotron~\cite{DBLP:conf/icassp/ValleLPC20} was to transfer the rhythm and pitch from reference audio to a target speaker via a Tacotron2-GST-based multi-speaker voice synthesis model, in which process a reference singing audio is necessary for the inference stage. To get rid of the dependency on the reference singing audio in the inference stage, Learn2Sing~1.0~\cite{DBLP:conf/slt/XueYLXL21} was proposed to train an SVS model for the target speaker with the help of a singing teacher, which makes the inference stage of the learned model like a typical SVS model that produces the singing voice with lyrics and musical scores as input. Specifically, in our previous Learn2Sing~1.0, independent duration and log-scale fundamental frequency (LF0) prediction models are trained to provide singing duration and pitch contours during the inference stage. Besides, considering the significant differences between the articulation patterns of speaking and singing, domain adversarial training~(DAT)~\cite{DBLP:journals/jmlr/GaninUAGLLML16} is employed to disentangle the style information from the acoustic features in the auto-regressive decoder. Most recently, another approach proposed in~\cite{DBLP:conf/asru/LiuWLSS21} using a similar framework to Learn2Sing~1.0 has been proposed, in which an F0 prediction module including vibrato modeling was used to generate natural singing voices.

However, it is challenging to explicitly predict the fundamental frequency of the singing voice. To be specific, an individual F0 prediction model accompanied by a limited amount of singing data may lead to inaccurate pitch curves~\cite{DBLP:journals/taslp/WangFYTW22}, and thus restrict the final performance of these F0 prediction-based methods~\cite{DBLP:conf/slt/XueYLXL21,DBLP:conf/asru/LiuWLSS21}. Furthermore, all above mentioned methods are based on the autoregressive decoder to predict the mel-spectrogram frame-by-frame, which makes the inference process slow. In addition, voice conversion is another solution to allow a speaker to sing without singing data~\cite{DBLP:conf/interspeech/ZhangYLWZWXLY20}, but this approach requires reference audio which makes it impractical to sing a new song, or to generate a highly expressive source speaker singing voice through an SVS system first, resulting in a lengthy pipeline.

To alleviate the above issues, a diffusion and mutual information-based SVS approach, referred to as Learn2Sing~2.0, is proposed to produce highly expressive singing for a target speaker by learning from the singing teacher. Learn2Sing~2.0 tries to synthesize an intermediate representation shared by different speakers, e.g., the speaker with or without singing corpus, and then restores the final waveform with additional style information, i.e, speaking or singing. In this way, the prediction of the intermediate representation can bridge the gap between the speech data and singing data regardless of the speaking style. To this end, inspired by the fact that pitch is the key style factor to distinguish singing from speaking voice, the intermediate representation is served by the mel-spectrogram with averaged pitch value in the phone level. Then, conditioned on the specific style, a diffusion decoder, which is accelerated by a fast sampling algorithm~\cite{DBLP:journals/corr/abs-2109-13821} during the inference stage, is adopted to gradually restore the final mel-spectrogram. As each speaker corresponds to a specific style, i.e., speaking or singing, it is necessary to make sure the speaker embedding is style-independent and vice versa. Here, Mutual Information~(MI)~\cite{DBLP:conf/icml/ChengHDLGC20} is employed to achieve the disentanglement of speaker and style. The generated mel-spectrogram is transformed to the final waveform by RefineGAN~\cite{DBLP:journals/corr/abs-2111-00962} which is designed for the SVS task. Experiments show that Learn2Sing~2.0 is capable of synthesizing high-quality singing voice for the target speaker without singing data with only 10 decoding steps. Moreover, the ablation study indicates the effectiveness of each component and the good design of Learn2Sing~2.0.

\vspace{-0.2cm}
\section{Proposed approach}

\begin{figure}[ht]
  \centering
  \includegraphics[width=8.5cm,height=9cm]{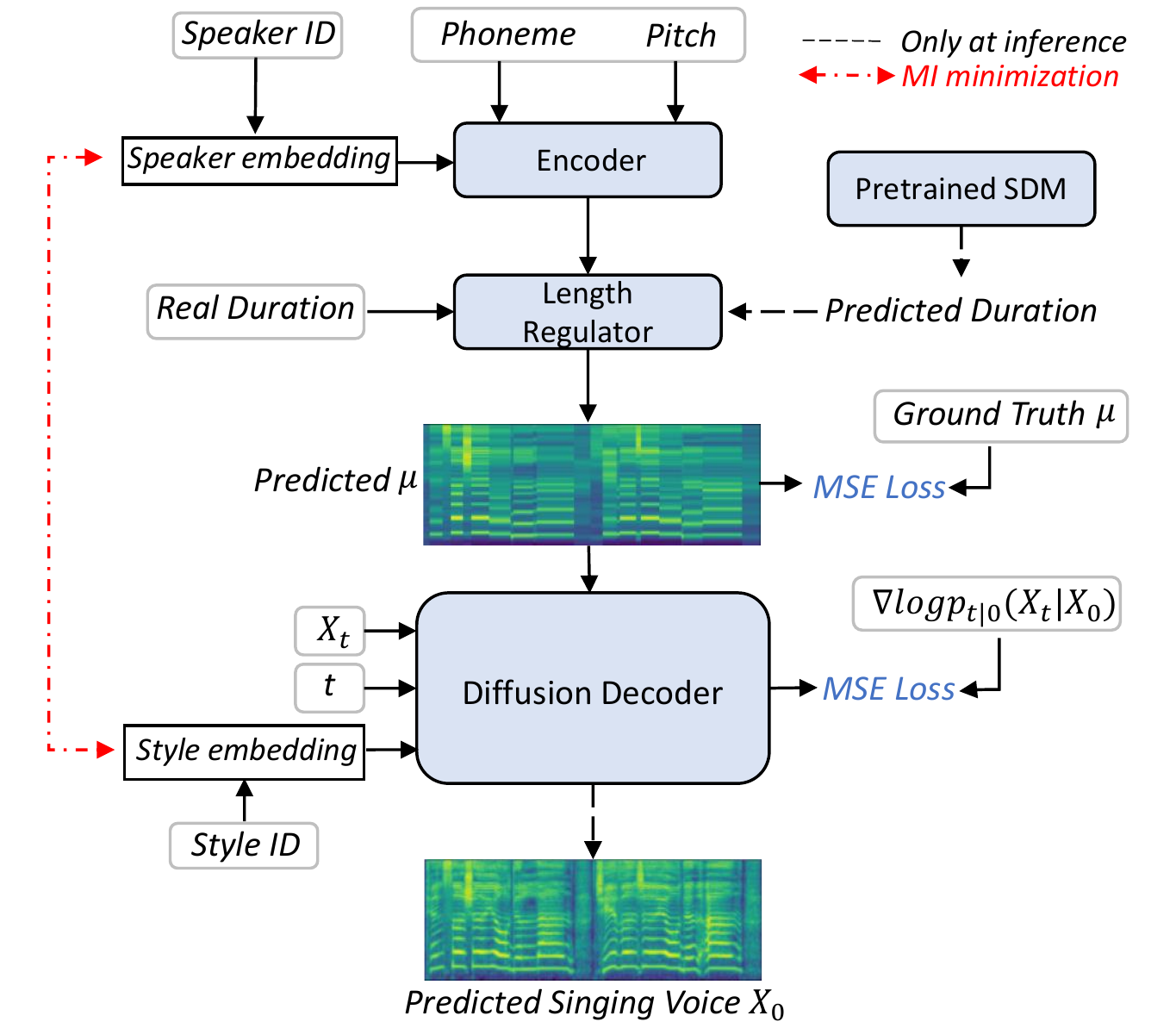}
  \caption{The training and inference process of the proposed approach.}
  \label{fig:speech_production}
\end{figure}

The proposed Learn2Sing~2.0 mainly consists of two parts. As shown in Fig.~\ref{fig:speech_production}, one is the generation of intermediate representations that are obtained by averaging the mel-spectrogram in phone level, which process takes the phone sequence, pitch, and speaker information as input, resulting in the features that are then up-sampled to align with the averaged mel-spectrogram. The second part is the restoration of the real mel-spectrogram based on the intermediate features and additional style information. This restoration stage is performed by a diffusion probabilistic model (DPM)-based decoder, which refers to Grad-TTS~\cite{DBLP:conf/icml/PopovVGSK21}. This section will first give a brief introduction to DPM, and then describe the proposed Learn2Sing~2 in detail.
\vspace{-0.2cm}
\subsection{Diffusion Probabilistic Model in TTS}
\vspace{-0.1cm}
To deal with the over-smoothing issue in mel-spectrogram generation caused by the MSE/MAE loss~\cite{DBLP:journals/corr/abs-2202-13066}, DPM is introduced into speech synthesis and SVS tasks due to its powerful modeling capabilities~\cite{DBLP:conf/interspeech/JeongKCCK21, liu2021diffsinger, DBLP:conf/icml/PopovVGSK21}. DPM generally has two processes, i.e, forward diffusion process and reverse diffusion process. The forward diffusion process obtains the tractable prior distribution by gradually adding noise to the real data, while the reverse diffusion process reduces the prior distribution to that of the real data. By modeling the two processes as a solution to It\^{o} stochastic differential equations~(SDEs), Song et al. propose a unified score-based DPM framework~\cite{DBLP:conf/iclr/0011SKKEP21} which is then applied to the speech synthesis domain in Grad-TTS. The two diffusion processes in Grad-TTS satisfies the following SDEs:
\begin{equation}
\setlength{\abovedisplayskip}{3pt}
\setlength{\belowdisplayskip}{3pt}
d X_{t}=\frac{1}{2} \Sigma^{-1}\left(\mu-X_{t}\right) \beta_{t} d t+\sqrt{\beta_{t}} d W_{t}, t \in[0, 1]
\end{equation}
\begin{equation}
\setlength{\abovedisplayskip}{3pt}
\setlength{\belowdisplayskip}{3pt}
\begin{aligned}
&d X_{t}=\left(\frac{1}{2} \Sigma^{-1}\left(\mu-X_{t}\right)-\nabla \log p_{t}\left(X_{t}|X_{0}\right)\right) \beta_{t} d t\\
&+\sqrt{\beta_{t}} d \widetilde{W}_{t}, \quad t \in[0, 1]
\end{aligned}
\end{equation}
where ${W}_{t}$ and $\widetilde{W}_{t}$ are forward and reverse-time Brownian motion, $\beta_{t}$ is the noise schedule for perturbing data with infinite number of noise scales, $\mu$ and $\Sigma$ is the mean and diagonal covariance matrix of the terminal distribution, $\log p_{t}\left(X_{t}|X_{0}\right)$ is the log probability density function which is predicted by a learnable score function $s_{\theta}\left(X_{t}, \mu, t\right)$ during the inference stage. Instead of solving the reverse-time SDE directly, Grad-TTS uses an ordinary differential equation~(ODE) in the reverse process: 
\begin{equation}\label{ODE}
\setlength{\abovedisplayskip}{3pt}
\setlength{\belowdisplayskip}{3pt}
d X_{t}=\frac{1}{2}\left(\mu-X_{t}-s_{\theta}\left(X_{t}, \mu, t\right)\right) \beta_{t} d t
\end{equation}
In brief, Grad-TTS learns to estimate gradients of log-density of $X_{t}$~(noisy mel-spectrogram) given $X_{0}$~(real mel-spectrogram) and $\mu$~(mel-spectrogram predicted from text) in the training processes. During the inference stage, Grad-TTS first predicts a mel-spectrogram from text and then gradually reconstructs the target mel-spectrogram using the score predicted from $s_{\theta}\left(X_{t}, \mu, t\right)$ in adjustable iterations.
\vspace{-0.3cm}
\vspace{-0.3cm}
\vspace{-0.2cm}
\subsection{Proposed Learn2Sing~2.0 Model}
\vspace{-0.1cm}
As illustrated in Fig.~\ref{fig:speech_production}, the input phoneme, pitch, and speaker embedding are encoded by an encoder and then a length regulator is used to up-sample the obtained feature. Regarding pitch as the most critical style factor in our task, in the training phase, the encoder generates phoneme-level averaged mel-spectrogram $\mu$ regardless of the speaking style using phoneme, pitch and speaker embedding as inputs. Note that there is no singing or speaking pitch style, e.g. overshoot, vibrato, preparation and fine fluctuation, in the predicted $\mu$ because all frames within the same phoneme are the same.  Then, the diffusion decoder predicts the gradient of the probability density function using the predicted $\mu$, $X_{t}$, $t$, and style embedding as inputs. In the inference phase, the encoder first generates the predicted $\mu$ without pitch style for the target speaker. The phoneme duration information can be provided by a pre-trained singing duration model~(SDM) or manual labeling interval files. Then the diffusion decoder gradually reconstructs the pitch contours with the singing style by solving the ODE, and finally generates the target speaker's singing voice $X_{0}$.

While the speaker embedding and style embedding are obtained from the speaker ID and style ID respectively, the  one-to-one correspondence between the speaker and style makes the model tend to be confused by the speaker and style information. In other words, the speaker embedding of one speaker may not only convey the speaker information but also the style information because the style, i.e., speaking or singing, of this speaker is specific. To face this challenge, mutual information is employed to further decouple style and speaker information. Specifically, to estimate and minimize the mutual information between speaker and style embeddings, vCLUB~\cite{DBLP:conf/icml/ChengHDLGC20} which is an approach that requires only samples to estimate an upper bound on the mutual information is adopted. With the sample pairs of speaker and style embeddings $\left\{\left(\boldsymbol{spk}_{i}, \boldsymbol{sty}_{i}\right)\right\}_{i=1}^{N}$, the MI loss is:
\begin{equation}
\setlength{\abovedisplayskip}{3pt}
\setlength{\belowdisplayskip}{3pt}
\begin{aligned}
\mathrm{\mathcal{L}_{mi}}=\frac{1}{N^{2}} \sum_{i=1}^{N} \sum_{j=1}^{N}\left[\log q_{\theta}\left(\boldsymbol{sty}_{i} \mid \boldsymbol{spk}_{i}\right) 
-\log  q_{\theta}\left(\boldsymbol{sty}_{j} \mid \boldsymbol{spk}_{i}\right)\right]
\end{aligned}
\end{equation}
where $\boldsymbol{spk}$ and $\boldsymbol{sty}$ are speaker and style embeddings, $q_{\theta}$ is a variational approximation which makes vCLUB a reliable MI estimator. In the training stage, we first obtain a batch of samples $\left\{\left(\boldsymbol{spk}_{i}, \boldsymbol{sty}_{i}\right)\right\}_{i=1}^{N}$ from Learn2Sing model and update the $q_{\theta}\left(\boldsymbol{sty} \mid \boldsymbol{spk}\right)$ by maximizing the log-likelihood:
\begin{equation}
\setlength{\abovedisplayskip}{3pt}
\setlength{\belowdisplayskip}{3pt}
\mathcal{L}(\theta)=\frac{1}{N} \sum_{i=1}^{N} \log q_{\theta}\left(\boldsymbol{sty}_{i} \mid \boldsymbol{spk}_{i}\right)
\end{equation}
The final objective function for training Learn2Sing is described as:
\begin{equation}
\setlength{\abovedisplayskip}{0pt}
\setlength{\belowdisplayskip}{0pt}
\mathcal{L}_{total} = \mathcal{L}_{\mu} + \mathcal{L}_{diff} + \lambda\mathcal{L}_{mi}
\end{equation}
where $\mathcal{L}_{\mu}$ is the MSE loss between the predicted $\mu$ and ground-truth $\mu$ that is achieved by averaging the real mel-spectrogram within each phoneme, $\mathcal{L}_{diff}$ is the diffusion loss, and $\lambda$ is the weight for $\mathcal{L}_{mi}$.
\vspace{-0.05cm}
\subsection{Fast Maximum Likelihood Sampling Scheme}
\vspace{-0.1cm}
For the diffusion-based decoder, adopting fewer inference steps is an effective method to speed up the inference process. However, reducing the decoding steps directly could degrade the quality of synthesized voice. In order to maintain the quality of the generated singing voice even with a very small number of steps, we introduce a fast sampling scheme proposed in~\cite{DBLP:journals/corr/abs-2109-13821}. This fast sampling scheme adopts a fixed-step first order reverse SDE solver which is designed to maximize the likelihood of discrete sample paths of the forward diffusion.  Specifically, the fast sampling scheme first defines the following values:
\begin{equation}
\setlength{\abovedisplayskip}{3pt}
\setlength{\belowdisplayskip}{3pt}
\begin{gathered}
\gamma_{s, t}=\exp \left(-\frac{1}{2} \int_{s}^{t} \beta_{u} d u\right),\phi_{s, t}=\gamma_{s, t} \frac{1-\gamma_{0,s}^{2}}{1-\gamma_{0, t}^{2}}, \\
\quad \nu_{s, t}=\gamma_{0, s} \frac{1-\gamma_{s, t}^{2}}{1-\gamma_{0, t}^{2}},\sigma_{s, t}^{2}=\frac{\left(1-\gamma_{0, s}^{2}\right)\left(1-\gamma_{s, t}^{2}\right)}{1-\gamma_{0, t}^{2}},\\
\kappa_{t, h}=\frac{\nu_{t-h, t}\left(1-\gamma_{0, t}^{2}\right)}{\gamma_{0, t} \beta_{t} h}-1, \quad\\
\omega_{t, h}=\frac{\phi_{t-h, t}-1}{\beta_{t} h}+\frac{1+\kappa_{t, h}}{1-\gamma_{0, t}^{2}}-\frac{1}{2}, \left(\sigma_{t, h}\right)^{2}=\sigma_{t-h, t}^{2}
\end{gathered}
\end{equation}
and adopts the following class of reverse SDE solvers:
\begin{equation}
\setlength{\abovedisplayskip}{3pt}
\setlength{\belowdisplayskip}{3pt}
\begin{aligned}
X_{t-h}=X_{t}+\beta_{t} h\left(\left(\frac{1}{2}+\omega_{t, h}\right)\left(X_{t}-\mu\right)
\right.\\
\phantom{=\;\;}
\left.
+\left(1+\kappa_{t, h}\right)s_{\theta}\left(X{t}, \mu, \boldsymbol{sty}, t\right)\right)+\sigma_{t, h} \xi_{t}
\end{aligned}
\end{equation}
where $0 \leq s<t \leq 1$, $h$ is the step size, $\xi_{t}$ are i.i.d. sampling from $\mathcal{N}(0, \mathrm{I})$,  and the derivation can be found in~\cite{DBLP:journals/corr/abs-2109-13821}.

\section{Experiments}
\vspace{-0.1cm}
\subsection{Data setups}
\vspace{-0.1cm}
Experiments to evaluate the performance of Learn2Sing~2.0 are performed on an internal corpus, which consists of three speakers, among which one is the teacher that provides the singing data and the other two speakers are students only having data. To be specific, the singing data recorded by the female singer contains 100 songs with around 5 hours, and each song is labeled with the music scores and the duration of each phoneme. As for the speakers working as students, one is from the open-source female TTS corpus Databaker-DB1\footnote{\url{Available at: www.data-baker.com/open_source.html}} which is referred to as Student-1, and the other one is from an internal female corpus referred as Student-2. All audio samples are downsampled to 22,050Hz and are then represented as 80-dimensional log-scale mel-spectrogram with 256 hop length and 1024 Hanning window. 
\vspace{-0.05cm}
\subsection{Model and hyper-parameters details}
\vspace{-0.1cm}
The encoder is identical to the encoder proposed in~\cite{DBLP:conf/nips/KimKKY20} to encode content, pitch and speaker information. The decoder follows that in~\cite{DBLP:conf/icml/PopovVGSK21} which is based on the UNet architecture. The maximum input frame length for the decoder is limited to 384 for more efficient use of memory. The dimensions of both style and speaker embedding are 128. The variational approximation $q_{\theta}\left(\boldsymbol{sty} \mid \boldsymbol{spk}\right)$ follows~\cite{DBLP:conf/interspeech/WangDYCLM21} which is parameterized in Gaussian distribution with mean and variance predicted by two linear layers. Besides, the weight for $\mathcal{L}_{mi}$ is set to 0.01 to achieve a stable disentanglement performance. With the help of the fast sampling scheme, the Learn2Sing 2.0 allows using fewer decoding steps to obtain high-quality results. Here, 10 decoding steps are used in the inference stage.  

To convert the mel-spectrogram into waveform, we trained a multi-speaker RefineGAN~\cite{DBLP:journals/corr/abs-2111-00962} using all the data. Since RefineGAN requires F0 as input to generate speech templates, but F0 is not explicitly predicted in our approach, we additionally trained a model, referred to as Mel2F0 which consists of 3 LSTM layers, to predict F0 from the mel-spectrogram. PYIN~\cite{DBLP:conf/icassp/MauchD14} is adopted to estimate F0 from the waveform. 

We reserved 4 complete songs from the singing corpus as the test set, while the rest of the singing data was used as the training and validation set. To avoid the effect of different duration models in the testing phase, the real phoneme level duration is used for different models. 
\vspace{-0.05cm}
\subsection{Evaluation on proposed Learn2Sing~2.0}
\vspace{-0.1cm}
Our task is evaluated in terms of the speech quality and speaker similarity of the synthesized singing voice. To be specific, a good learn2sing model should create the singing voice not only with high quality but also with the timbre of the target (student) speaker. Here, the opinion score (MOS) tests are conducted to subjectively evaluate the synthesized results in terms of naturalness and similarity. Besides, the F0 Mean Absolute Error~(MAE) is also measured to objectively evaluate the pitch accuracy of the synthesized singing voice. Furthermore, the inference speed is also assessed in terms of the average real time factor~(RTF) on GPU. 

\setlength{\tabcolsep}{4mm}{
\begin{table*}[htbp]
  \centering
  \caption{Experiment results of MOS with 95\% confidence intervals, F0 MAE and RTF}
    \begin{tabular}{cccccc}
    \toprule
    Model & Target Speaker & Naturalness & Similarity & F0 MAE & RTF \\
    \midrule
    \multirow{3}[2]{*}{Learn2Sing 1.0}
     & Teacher & 3.85±0.15 &   -    & 3.33  &  \\
    & Student-1 & 3.20±0.09 & 3.14±0.08 & 16.27 & 0.098\multirow{3}[2]{*}{} \\
          & Student-2 & 3.28±0.10 & 3.30±0.07 & 15.67 &  \\
         
    \midrule
    \multirow{3}[2]{*}{Learn2Sing 2.0} 
      & Teacher & 4.04±0.14 &   -    & 11.12 &  \\
    & Student-1 & 3.41±0.12 & 3.79±0.08 & 15.49 & 0.050\multirow{3}[2]{*}{} \\
          & Student-2 & 3.43±0.08 & 3.84±0.07 & 14.3  &  \\
        
    \midrule
    Recording &       & 4.50±0.07 &   -    &   -    &  \\
    \bottomrule
    \end{tabular}%
  \label{tab:mos}%
\end{table*}%
}

The results are shown in Table~\ref{tab:mos}, in which the Learn2Sing~1.0 is compared. For Learn2sing~1.0, the ground-truth phoneme duration and frame level F0 are adopted in the inference phase. In this table, the target speaker with \textit{Teacher} indicates a general SVS task rather than the learn2sing task, which is to present the singing voice synthesis capability of different models and also provides an upper bound for a model in the learn2sing task. As can be seen from this table, the MOS values in this general SVS task are 3.85 and 4.0 achieved by Learn2Sing~1.0 and Learn2Sing~2.0, respectively, which indicates both approaches are capable of conducting singing voice synthesis well. Learn2Sing~1.0 outperforms Learn2Sing~2.0 in terms of F0 MAE due to the use of the real F0 curve as inputs in Learn2Sing~1.0.

In the learn2sing task, the synthesized singing samples of different models for both students are inferior to the samples generated via the general SVS in terms of all evaluation metrics. This phenomenon is straightforward that the absence of singing data for the target speaker makes the learn2sing a more challenging task compared with the general SVS task. Compared with Learn2Sing~1.0, the proposed Learn2Sing~2.0 achieves obvious superiority in terms of all evaluation metrics, especially the speaker similarity in which the MOS values obtained by the proposed method are 20.7~\% and 16.4~\% relatively higher than that achieved by Learn2Sing~1.0 for Student-1 and Student-2 respectively. As for the inference speed, the RTF of Learn2Sing~2.0 is only 51.0~\% of Learn2Sing~1.0, demonstrating the fast inference speed of Learn2Speech~2.0, which is attributed to the fast sampling scheme that allows us to use fewer decoding steps in the inference stage.
\vspace{-0.2cm}
\subsection{Ablation study}
\vspace{-0.2cm}
The results are shown in Table~\ref{MOS for ablation study}, in which the MOS values of both naturalness and similarity are reported on the Student-1. It is important to highlight that calculating the $\mu$ loss with the real mel-spectrogram rather than the averaged  mel-spectrogram results in the worst performance, which demonstrates the practicality of the pipeline that restores the final mel-spectrogram based on the intermediate representation with pitch averaged at phone level. 

\setlength{\tabcolsep}{3mm}{
\begin{table}[b]
	\centering
	\caption{Ablation study with MOS test. w/o means without. w/o phone average $\mu$ means that the ground-truth $\mu$ is real mel-spectrogram rather than the mel-spectrogram averaged in the phone level.}
	\begin{tabular}{lll}
		\toprule
		Model & Naturalness&Similarity  \\ 
		\midrule
		Learn2Sing 2.0-FS-10                &\textbf{3.41±0.12}  &\textbf{3.79±0.08}  \\
        \ w/o~Mutual information  & 3.20±0.12 &3.52±0.08 \\
        \ w/o~Fast sampling       & 3.07±0.12 & 3.46±0.09 \\
        \  w/o~$\mu$ Loss &3.37±0.12  &3.52±0.07\\
        \  w/o~phone average $\mu$ & 3.01±0.13 & 3.45±0.09 \\
		\bottomrule
	\end{tabular}
	\label{MOS for ablation study}
\end{table}
}

To reveal why listeners prefer the proposed approach, we further visualized the generated mel-spectrograms for several test clips from different models, and the results are presented in Fig.~2. Comparing the mel-spectrogram in the blue box, only the proposed Learn2Sing 2.0 does not include glitches, suggesting that the mel-spectrogram generated by the proposed approach is closer to the real mel-spectrogram. The green box contains melisma\footnote{In singing, the term melisma refers to a passage of music that has a group of notes that are sung with just one syllable of text.} skill in singing, and it can be seen that Learn2Sing 2.0 has more natural melisma.\footnote{We recommend that listeners listen to the full test set of songs generated by different models from:\url{https://welkinyang.github.io/Learn2Sing2.0/}}
\vspace{-0.3cm}
\begin{figure}[H]
  \centering
  \includegraphics[width=7cm,height=5.5cm]{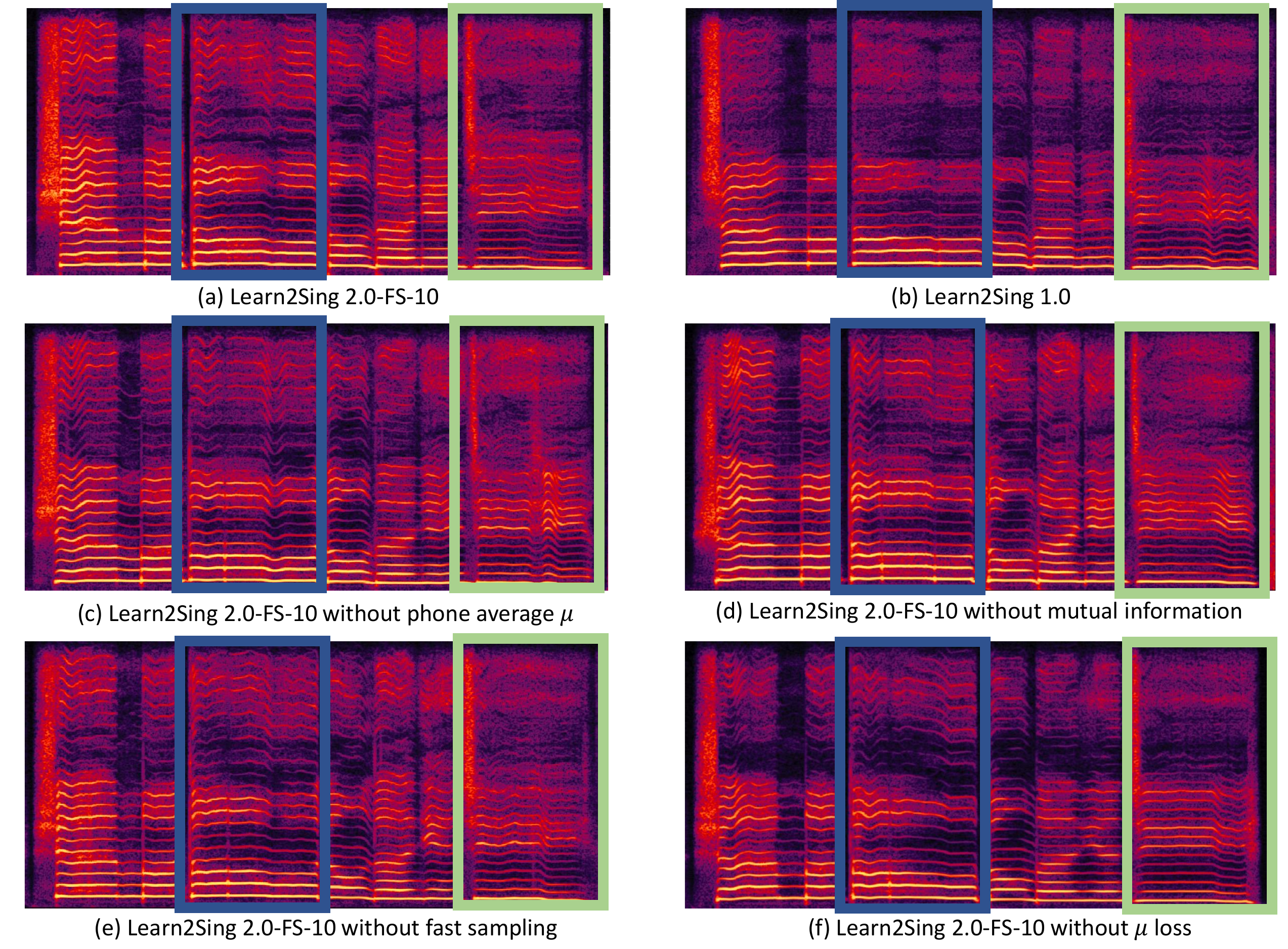}
  \caption{Mel-spectrogram of a testing sample for different models.}
  \label{fig:mel_compare}
\end{figure}
\vspace{-0.7cm}
\section{Conclusions}
\vspace{-0.1cm}
This paper presents a novel approach, referred to as Learn2Sing~2.0, to generate the singing voice for a target speaker with only speech data with the help of the singing teacher. Learn2Sing~2.0 first generates the preliminary acoustic feature with averaged pitch value in the phone level. Then, with additional style information, i.e., speaking or singing, a diffusion decoder is adopted to restore the final mel-spectrogram. Mutual information is introduced in Learn2Sing~2.0 to disentangle the style and speaker information. Besides, to speed up the inference speed, a fast sampling algorithm for the diffusion-based decoder is introduced in the inference stage. Extensive experiments demonstrate that the proposed approach is capable of synthesizing high-quality singing voice for a target speaker without any singing data at a high inference speed. Moreover, the ablation study indicates the effectiveness of each component and the good design of Learn2Sing~2.0.
\bibliographystyle{IEEEtran}

\bibliography{mybib}

\begin{thebibliography}{10}
\providecommand{\url}[1]{#1}
\csname url@samestyle\endcsname
\providecommand{\newblock}{\relax}
\providecommand{\bibinfo}[2]{#2}
\providecommand{\BIBentrySTDinterwordspacing}{\spaceskip=0pt\relax}
\providecommand{\BIBentryALTinterwordstretchfactor}{4}
\providecommand{\BIBentryALTinterwordspacing}{\spaceskip=\fontdimen2\font plus
\BIBentryALTinterwordstretchfactor\fontdimen3\font minus
  \fontdimen4\font\relax}
\providecommand{\BIBforeignlanguage}[2]{{%
\expandafter\ifx\csname l@#1\endcsname\relax
\typeout{** WARNING: IEEEtran.bst: No hyphenation pattern has been}%
\typeout{** loaded for the language `#1'. Using the pattern for}%
\typeout{** the default language instead.}%
\else
\language=\csname l@#1\endcsname
\fi
#2}}
\providecommand{\BIBdecl}{\relax}
\BIBdecl

\bibitem{DBLP:conf/iscslp/GuYRWTZCWM21}
\BIBentryALTinterwordspacing
Y.~Gu, X.~Yin, Y.~Rao, Y.~Wan, B.~Tang, Y.~Zhang, J.~Chen, Y.~Wang, and Z.~Ma,
  ``Bytesing: {A} chinese singing voice synthesis system using duration
  allocated encoder-decoder acoustic models and wavernn vocoders,'' in
  \emph{12th International Symposium on Chinese Spoken Language Processing,
  {ISCSLP} 2021, Hong Kong, January 24-27, 2021}.\hskip 1em plus 0.5em minus
  0.4em\relax {IEEE}, 2021, pp. 1--5. [Online]. Available:
  \url{https://doi.org/10.1109/ISCSLP49672.2021.9362104}
\BIBentrySTDinterwordspacing

\bibitem{DBLP:conf/interspeech/LeeKBLKC21}
\BIBentryALTinterwordspacing
G.~Lee, T.~Kim, H.~Bae, M.~Lee, Y.~Kim, and H.~Cho, ``N-singer: {A}
  non-autoregressive korean singing voice synthesis system for pronunciation
  enhancement,'' in \emph{Interspeech 2021, 22nd Annual Conference of the
  International Speech Communication Association, Brno, Czechia, 30 August - 3
  September 2021}, H.~Hermansky, H.~Cernock{\'{y}}, L.~Burget, L.~Lamel,
  O.~Scharenborg, and P.~Motl{\'{\i}}cek, Eds.\hskip 1em plus 0.5em minus
  0.4em\relax {ISCA}, 2021, pp. 1589--1593. [Online]. Available:
  \url{https://doi.org/10.21437/Interspeech.2021-239}
\BIBentrySTDinterwordspacing

\bibitem{DBLP:journals/corr/abs-2110-08813}
\BIBentryALTinterwordspacing
Y.~Zhang, J.~Cong, H.~Xue, L.~Xie, P.~Zhu, and M.~Bi, ``Visinger: Variational
  inference with adversarial learning for end-to-end singing voice synthesis,''
  \emph{CoRR}, vol. abs/2110.08813, 2021. [Online]. Available:
  \url{https://arxiv.org/abs/2110.08813}
\BIBentrySTDinterwordspacing

\bibitem{DBLP:journals/corr/abs-2201-07429}
\BIBentryALTinterwordspacing
Y.~Wang, X.~Wang, P.~Zhu, J.~Wu, H.~Li, H.~Xue, Y.~Zhang, L.~Xie, and M.~Bi,
  ``Opencpop: {A} high-quality open source chinese popular song corpus for
  singing voice synthesis,'' \emph{CoRR}, vol. abs/2201.07429, 2022. [Online].
  Available: \url{https://arxiv.org/abs/2201.07429}
\BIBentrySTDinterwordspacing

\bibitem{DBLP:conf/mm/HuangC0LCZ21}
\BIBentryALTinterwordspacing
R.~Huang, F.~Chen, Y.~Ren, J.~Liu, C.~Cui, and Z.~Zhao, ``Multi-singer: Fast
  multi-singer singing voice vocoder with {A} large-scale corpus,'' in
  \emph{{MM} '21: {ACM} Multimedia Conference, Virtual Event, China, October 20
  - 24, 2021}, H.~T. Shen, Y.~Zhuang, J.~R. Smith, Y.~Yang, P.~Cesar, F.~Metze,
  and B.~Prabhakaran, Eds.\hskip 1em plus 0.5em minus 0.4em\relax {ACM}, 2021,
  pp. 3945--3954. [Online]. Available:
  \url{https://doi.org/10.1145/3474085.3475437}
\BIBentrySTDinterwordspacing

\bibitem{DBLP:journals/speech/SharmaGVTL21}
\BIBentryALTinterwordspacing
B.~Sharma, X.~Gao, K.~Vijayan, X.~Tian, and H.~Li, ``{NHSS:} {A} speech and
  singing parallel database,'' \emph{Speech Commun.}, vol. 133, pp. 9--22,
  2021. [Online]. Available: \url{https://doi.org/10.1016/j.specom.2021.07.002}
\BIBentrySTDinterwordspacing

\bibitem{DBLP:conf/icassp/ValleLPC20}
\BIBentryALTinterwordspacing
R.~Valle, J.~Li, R.~Prenger, and B.~Catanzaro, ``Mellotron: Multispeaker
  expressive voice synthesis by conditioning on rhythm, pitch and global style
  tokens,'' in \emph{2020 {IEEE} International Conference on Acoustics, Speech
  and Signal Processing, {ICASSP} 2020, Barcelona, Spain, May 4-8, 2020}.\hskip
  1em plus 0.5em minus 0.4em\relax {IEEE}, 2020, pp. 6189--6193. [Online].
  Available: \url{https://doi.org/10.1109/ICASSP40776.2020.9054556}
\BIBentrySTDinterwordspacing

\bibitem{DBLP:conf/slt/XueYLXL21}
\BIBentryALTinterwordspacing
H.~Xue, S.~Yang, Y.~Lei, L.~Xie, and X.~Li, ``Learn2sing: Target speaker
  singing voice synthesis by learning from a singing teacher,'' in \emph{{IEEE}
  Spoken Language Technology Workshop, {SLT} 2021, Shenzhen, China, January
  19-22, 2021}.\hskip 1em plus 0.5em minus 0.4em\relax {IEEE}, 2021, pp.
  522--529. [Online]. Available:
  \url{https://doi.org/10.1109/SLT48900.2021.9383585}
\BIBentrySTDinterwordspacing

\bibitem{DBLP:conf/asru/LiuWLSS21}
\BIBentryALTinterwordspacing
R.~Liu, X.~Wen, C.~Lu, L.~Song, and J.~S. Sung, ``Vibrato learning in
  multi-singer singing voice synthesis,'' in \emph{{IEEE} Automatic Speech
  Recognition and Understanding Workshop, {ASRU} 2021, Cartagena, Colombia,
  December 13-17, 2021}.\hskip 1em plus 0.5em minus 0.4em\relax {IEEE}, 2021,
  pp. 773--779. [Online]. Available:
  \url{https://doi.org/10.1109/ASRU51503.2021.9688029}
\BIBentrySTDinterwordspacing

\bibitem{DBLP:journals/jmlr/GaninUAGLLML16}
\BIBentryALTinterwordspacing
Y.~Ganin, E.~Ustinova, H.~Ajakan, P.~Germain, H.~Larochelle, F.~Laviolette,
  M.~Marchand, and V.~S. Lempitsky, ``Domain-adversarial training of neural
  networks,'' \emph{J. Mach. Learn. Res.}, vol.~17, pp. 59:1--59:35, 2016.
  [Online]. Available: \url{http://jmlr.org/papers/v17/15-239.html}
\BIBentrySTDinterwordspacing

\bibitem{DBLP:journals/taslp/WangFYTW22}
\BIBentryALTinterwordspacing
T.~Wang, R.~Fu, J.~Yi, J.~Tao, and Z.~Wen, ``Neuraldps: Neural deterministic
  plus stochastic model with multiband excitation for noise-controllable
  waveform generation,'' \emph{{IEEE} {ACM} Trans. Audio Speech Lang.
  Process.}, vol.~30, pp. 865--878, 2022. [Online]. Available:
  \url{https://doi.org/10.1109/TASLP.2022.3140480}
\BIBentrySTDinterwordspacing

\bibitem{DBLP:conf/interspeech/ZhangYLWZWXLY20}
\BIBentryALTinterwordspacing
L.~Zhang, C.~Yu, H.~Lu, C.~Weng, C.~Zhang, Y.~Wu, X.~Xie, Z.~Li, and D.~Yu,
  ``Durian-sc: Duration informed attention network based singing voice
  conversion system,'' in \emph{Interspeech 2020, 21st Annual Conference of the
  International Speech Communication Association, Virtual Event, Shanghai,
  China, 25-29 October 2020}, H.~Meng, B.~Xu, and T.~F. Zheng, Eds.\hskip 1em
  plus 0.5em minus 0.4em\relax {ISCA}, 2020, pp. 1231--1235. [Online].
  Available: \url{https://doi.org/10.21437/Interspeech.2020-1789}
\BIBentrySTDinterwordspacing

\bibitem{DBLP:journals/corr/abs-2109-13821}
\BIBentryALTinterwordspacing
V.~Popov, I.~Vovk, V.~Gogoryan, T.~Sadekova, M.~A. Kudinov, and J.~Wei,
  ``Diffusion-based voice conversion with fast maximum likelihood sampling
  scheme,'' \emph{CoRR}, vol. abs/2109.13821, 2021. [Online]. Available:
  \url{https://arxiv.org/abs/2109.13821}
\BIBentrySTDinterwordspacing

\bibitem{DBLP:conf/icml/ChengHDLGC20}
\BIBentryALTinterwordspacing
P.~Cheng, W.~Hao, S.~Dai, J.~Liu, Z.~Gan, and L.~Carin, ``{CLUB:} {A}
  contrastive log-ratio upper bound of mutual information,'' in
  \emph{Proceedings of the 37th International Conference on Machine Learning,
  {ICML} 2020, 13-18 July 2020, Virtual Event}, ser. Proceedings of Machine
  Learning Research, vol. 119.\hskip 1em plus 0.5em minus 0.4em\relax {PMLR},
  2020, pp. 1779--1788. [Online]. Available:
  \url{http://proceedings.mlr.press/v119/cheng20b.html}
\BIBentrySTDinterwordspacing

\bibitem{DBLP:journals/corr/abs-2111-00962}
\BIBentryALTinterwordspacing
S.~Xu, W.~Zhao, and J.~Guo, ``Refinegan: Universally generating waveform better
  than ground truth with highly accurate pitch and intensity responses,''
  \emph{CoRR}, vol. abs/2111.00962, 2021. [Online]. Available:
  \url{https://arxiv.org/abs/2111.00962}
\BIBentrySTDinterwordspacing

\bibitem{DBLP:conf/icml/PopovVGSK21}
\BIBentryALTinterwordspacing
V.~Popov, I.~Vovk, V.~Gogoryan, T.~Sadekova, and M.~A. Kudinov, ``Grad-tts: {A}
  diffusion probabilistic model for text-to-speech,'' in \emph{Proceedings of
  the 38th International Conference on Machine Learning, {ICML} 2021, 18-24
  July 2021, Virtual Event}, ser. Proceedings of Machine Learning Research,
  M.~Meila and T.~Zhang, Eds., vol. 139.\hskip 1em plus 0.5em minus 0.4em\relax
  {PMLR}, 2021, pp. 8599--8608. [Online]. Available:
  \url{http://proceedings.mlr.press/v139/popov21a.html}
\BIBentrySTDinterwordspacing

\bibitem{DBLP:journals/corr/abs-2202-13066}
\BIBentryALTinterwordspacing
Y.~Ren, X.~Tan, T.~Qin, Z.~Zhao, and T.~Liu, ``Revisiting over-smoothness in
  text to speech,'' \emph{CoRR}, vol. abs/2202.13066, 2022. [Online].
  Available: \url{https://arxiv.org/abs/2202.13066}
\BIBentrySTDinterwordspacing

\bibitem{DBLP:conf/interspeech/JeongKCCK21}
\BIBentryALTinterwordspacing
M.~Jeong, H.~Kim, S.~J. Cheon, B.~J. Choi, and N.~S. Kim, ``Diff-tts: {A}
  denoising diffusion model for text-to-speech,'' in \emph{Interspeech 2021,
  22nd Annual Conference of the International Speech Communication Association,
  Brno, Czechia, 30 August - 3 September 2021}, H.~Hermansky,
  H.~Cernock{\'{y}}, L.~Burget, L.~Lamel, O.~Scharenborg, and
  P.~Motl{\'{\i}}cek, Eds.\hskip 1em plus 0.5em minus 0.4em\relax {ISCA}, 2021,
  pp. 3605--3609. [Online]. Available:
  \url{https://doi.org/10.21437/Interspeech.2021-469}
\BIBentrySTDinterwordspacing

\bibitem{liu2021diffsinger}
J.~Liu, C.~Li, Y.~Ren, F.~Chen, P.~Liu, and Z.~Zhao, ``Diffsinger: Singing
  voice synthesis via shallow diffusion mechanism,'' \emph{arXiv preprint
  arXiv:2105.02446}, vol.~2, 2021.

\bibitem{DBLP:conf/iclr/0011SKKEP21}
\BIBentryALTinterwordspacing
Y.~Song, J.~Sohl{-}Dickstein, D.~P. Kingma, A.~Kumar, S.~Ermon, and B.~Poole,
  ``Score-based generative modeling through stochastic differential
  equations,'' in \emph{9th International Conference on Learning
  Representations, {ICLR} 2021, Virtual Event, Austria, May 3-7, 2021}.\hskip
  1em plus 0.5em minus 0.4em\relax OpenReview.net, 2021. [Online]. Available:
  \url{https://openreview.net/forum?id=PxTIG12RRHS}
\BIBentrySTDinterwordspacing

\bibitem{DBLP:conf/nips/KimKKY20}
\BIBentryALTinterwordspacing
J.~Kim, S.~Kim, J.~Kong, and S.~Yoon, ``Glow-tts: {A} generative flow for
  text-to-speech via monotonic alignment search,'' in \emph{Advances in Neural
  Information Processing Systems 33: Annual Conference on Neural Information
  Processing Systems 2020, NeurIPS 2020, December 6-12, 2020, virtual},
  H.~Larochelle, M.~Ranzato, R.~Hadsell, M.~Balcan, and H.~Lin, Eds., 2020.
  [Online]. Available:
  \url{https://proceedings.neurips.cc/paper/2020/hash/5c3b99e8f92532e5ad1556e53ceea00c-Abstract.html}
\BIBentrySTDinterwordspacing

\bibitem{DBLP:conf/interspeech/WangDYCLM21}
\BIBentryALTinterwordspacing
D.~Wang, L.~Deng, Y.~T. Yeung, X.~Chen, X.~Liu, and H.~Meng, ``{VQMIVC:} vector
  quantization and mutual information-based unsupervised speech representation
  disentanglement for one-shot voice conversion,'' in \emph{Interspeech 2021,
  22nd Annual Conference of the International Speech Communication Association,
  Brno, Czechia, 30 August - 3 September 2021}, H.~Hermansky,
  H.~Cernock{\'{y}}, L.~Burget, L.~Lamel, O.~Scharenborg, and
  P.~Motl{\'{\i}}cek, Eds.\hskip 1em plus 0.5em minus 0.4em\relax {ISCA}, 2021,
  pp. 1344--1348. [Online]. Available:
  \url{https://doi.org/10.21437/Interspeech.2021-283}
\BIBentrySTDinterwordspacing

\bibitem{DBLP:conf/icassp/MauchD14}
\BIBentryALTinterwordspacing
M.~Mauch and S.~Dixon, ``{PYIN:} {A} fundamental frequency estimator using
  probabilistic threshold distributions,'' in \emph{{IEEE} International
  Conference on Acoustics, Speech and Signal Processing, {ICASSP} 2014,
  Florence, Italy, May 4-9, 2014}.\hskip 1em plus 0.5em minus 0.4em\relax
  {IEEE}, 2014, pp. 659--663. [Online]. Available:
  \url{https://doi.org/10.1109/ICASSP.2014.6853678}
\BIBentrySTDinterwordspacing

\end{thebibliography}

\end{document}